\documentclass[onecollarge,runningheads]{svjour2}
\smartqed  
\usepackage[numbers]{natbib}
\usepackage{graphicx}
\usepackage{subfig}
\journalname{Journal of Biological Physics}
\begin{document}

\title{DNA like$-$charge attraction and overcharging by divalent counterions
in the presence of divalent co$-$ions}

\titlerunning{Divalent co-ion effects on DNA like charge attraction and overcharging}

\author{Nguyen Viet Duc, Toan T. Nguyen, and Paolo Carloni }


\institute{Nguyen Viet Duc \at
	Faculty of Physics, VNU University of Science, Vietnam National University,\\
	334 Nguyen Trai Street, Thanh Xuan, Hanoi, Vietnam\\
        \and
        Toan T. Nguyen \at
              VNU Key Laboratory ``Multiscale Simulation of Complex Systems'',\\
              VNU University of Science, Vietnam National University, \\
              334 Nguyen Trai Street, Thanh Xuan, Hanoi, Vietnam. \\
              \email{toannt@hus.edu.vn, toannt@vnu.edu.vn}\\
              Faculty of Physics, VNU University of Science, Vietnam National University,\\
              334 Nguyen Trai Street, Thanh Xuan, Hanoi, Vietnam\\
              School of Physics, Georgia Institute of Technology, 
              837 State Street, Atlanta, Georgia 30332-0430, USA\\
              \email{toan.nguyen@physics.gatech.edu}
        \and
       	Paolo Carloni \at
       	Computational Biomedicine, Institute for Advanced Simulation IAS-5, 
       	and Institute of Neuroscience and Medicine INM-9, Forschungszentrum Julich, 52425 Julich, Germany\\
}

\date{Received: date / Accepted: date}

\maketitle

\begin{abstract}
Strongly correlated electrostatics of DNA systems has drawn the 
interest of many groups, especially the condensation
and overcharging of DNA by multivalent counterions. 
By adding counterions of different valencies and shapes,
one can enhance or reduce DNA overcharging. In this papers, 
we focus on the effect of multivalent co-ions, specifically divalent
co-ions such as SO$_4^{2-}$. A computational experiment 
of DNA condensation using Monte$-$Carlo simulation in
grand canonical ensemble is carried out where DNA system is
in equilibrium with a bulk solution containing a mixture
of salt of different valency of co-ions.
Compared to system with purely monovalent co-ions,
the influence of divalent co-ions shows up in multiple aspects.
Divalent co-ions lead to an increase of 
monovalent salt in the DNA condensate. 
Because monovalent salts mostly
participate in linear screening of electrostatic interactions in the system,
more monovalent salt molecules enter the condensate leads to screening out of
short-range DNA$-$DNA like charge attraction and weaker DNA condensation free energy.
The overcharging of DNA by multivalent counterions is also reduced
in the presence of divalent co$-$ions.
Strong repulsions between DNA and divalent co-ions
and among divalent co-ions themselves leads to a {\em depletion} of negative ions
near DNA surface as compared to the case without divalent
co-ions. At large distance, the DNA$-$DNA repulsive 
interaction is stronger in the presence of divalent co$-$ions,
suggesting that divalent co$-$ions role is not only 
that of simple stronger linear screening.
\end{abstract}

\section{Introduction}
\label{intro}
The problem of DNA condensation in the presence of multivalent counterions
has seen a strong revival of interest in recent years. 
DNA study enables us to find effective ways of
gene delivery for the rapidly growing field of genetic therapy. 
Condensation of DNA inside viruses such as bacteriophages
provides excellent study candidates for this purpose
(see review \cite{GelbartVirusReview2009}). Furthermore, 
The promising development of DNA-based nanotechnology that can
be controlled exquisitely at nanoscale into precise 2D and 3D shapes
enables the fabrication of precise nanoscale devices 
that have already shown great potential for biomedical applications such as: 
drug delivery, biosensing, and synthetic nanopore formation
\cite{DNANNano2015,DNAScience2015}.

In aqueous solution, the phosphate group of DNA nucleotide becomes
negatively charged. Hence, DNA is charged negatively, with
the linear charge density of about 1e/1.7\AA, and surface charge density
of about 1$e$/1nm$^2$. These are among the highest charge densities for
biological molecules, therefore, electrostatics and the screening condition of the solution
play an important role in the structure and functions of DNA systems.
One of the most interesting and important electrostatics of DNA system
is the ability to condense DNA by multivalent counterions
by CoHex$^{+3}$, Spd$^{+3}$ or Spm$^{+4}$
\cite{Parsegian92,Hud01,HoangTorroidJCP2014,GrasonPRL2010}. 
In a restricted environment such as inside a viral capsid,
even divalent counterions such as Mg$^{+2}$ can also
cause strong and non-monotonic effect \cite{Knobler08}. 
By varying the salinity of solution, one can vary the 
amount of DNA ejected from viruses. 
There is an optimal counterion concentration, $c_{Z,0}$, where 
the least DNA genome is ejected from the phages. 
For counterion concentration, $c_Z$, higher or lower than this optimal concentration,
more DNA is ejected from phages. 

The non-monotonic influence of multivalent counterions on DNA ejection from viruses 
is expected to have the same physical origin as the phenomenon of
reentrant DNA condensation in free solution in the presence
of counterions of tri-, tetra- and higher valence
\cite{NguyenJCP2000,SaminathanBiochem1999,LivolantBJ1996,NetzLikeChargedRods,GelbartPhysToday}.
This can be understood as following.
At low concentration of the counterions, DNA is undercharged.
At high concentration of the counterions, DNA is overcharged (becomes
a positively charged molecule). Electrostatic repulsions
among DNA molecules keep them from condensation. 
At intermediate concentration, DNA is almost neutral thus, they
are able to condense into bundle under short range attraction forces.
Although, divalent counterions are known to condense
DNA only partially in free solution \cite{Parsegian92,Hud01}, DNA virus
provides a unique experimental setup. The constraint of
the viral capsid strongly eliminates configurational entropic cost of
packaging DNA. This allows divalent counterions to influence
DNA condensation similar to that of trivalent/tetravalent
counterions. Indeed, DNA condensation by divalent counterions has
also been observed in another environment where DNA configuration is 
constrained, namely the 
condensation of DNA in two dimensional systems \cite{Koltover2000}. 
For virus systems, theoretical fitting suggests that the DNA is neutralized 
at $c_{Z,0}\approx 75$mM for divalent counterions, and the short$-$range DNA 
attraction at this concentration is $-0.004k_BT$ per nucleotide 
base \cite{NguyenJCP2011,NguyenJBP2013}. 

The marginal case of divalent counterions also shows ion specificity effect.
For example, in free solution, Mn$^{+2}$ ions are able to 
condense DNA into disorder bundle but Mg$^{+2}$ ions are not \cite{Parsegian92}.
In DNA ejection from bacteriophages, the non-monotonicity is 
observed for MgSO$_4$ salt but not for MgCl$_2$ salt up to 
the concentration of 100mM \cite{Knobler08}. In a recent
paper \cite{NguyenJCP2016}, it is been shown that some of these
ion specificity effects can be captured by studying the difference
in the hydration radius of different counterions.
In this paper, we focus on the effect of divalent {\em co-ions} on the overcharging and condensation
of DNA by multivalent counterions. The problem is studied using 
a Monte-Carlo simulation of 
the DNA system in the grand canonical ensemble.
It is shown that divalent coions make it easier for monovalent salt
to enter the condensate to screen out and weaken DNA$-$DNA
interactions. At the same time, they cause a depletion of
negative ions near DNA surface which enhance DNA charge inversion
by multivalent counterions. At large
distance, the repulsive interaction among
undercharged or overcharged DNA is stronger in the presence
divalent co$-$ions. Some preliminiary result of this work
was presented in Ref. \cite{DucJPCS2016}.

The paper is organized as follow. In Sec. 2, the simulation model and 
method is presented. In Sec. 3, various simulation results
on the fugacities of the salts, the ion concentration variation
in DNA bundles, the osmotic pressure and DNA packaging free energy
are presented and discussed. We conclude in Sec. 4.

\section{The simulation model and method}

\begin{figure}[htbp]
\centering
\hfill
\subfloat[]{\includegraphics[height=2.5in]{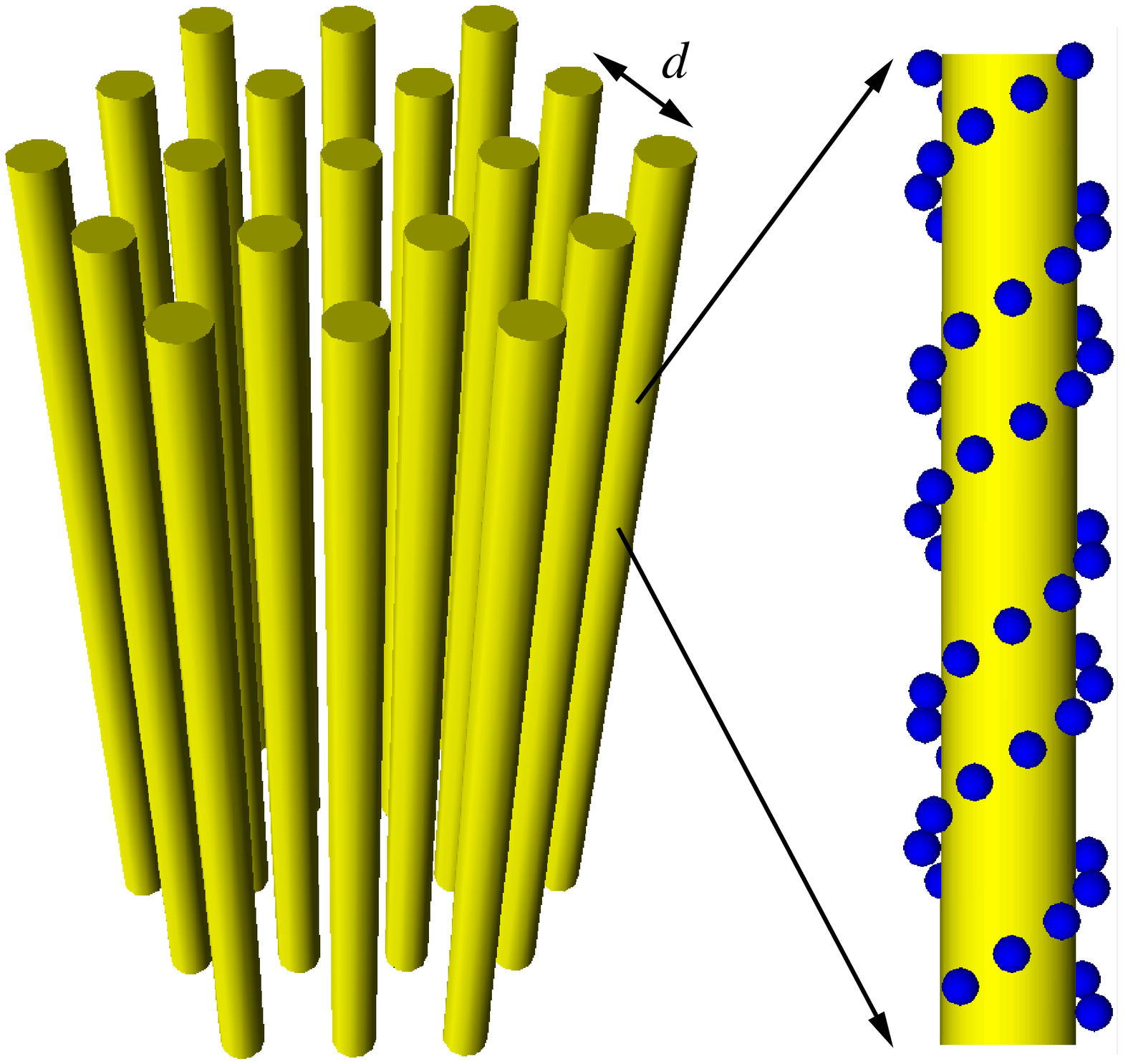}}
\hfill
\subfloat[]{\includegraphics[height=2.3in]{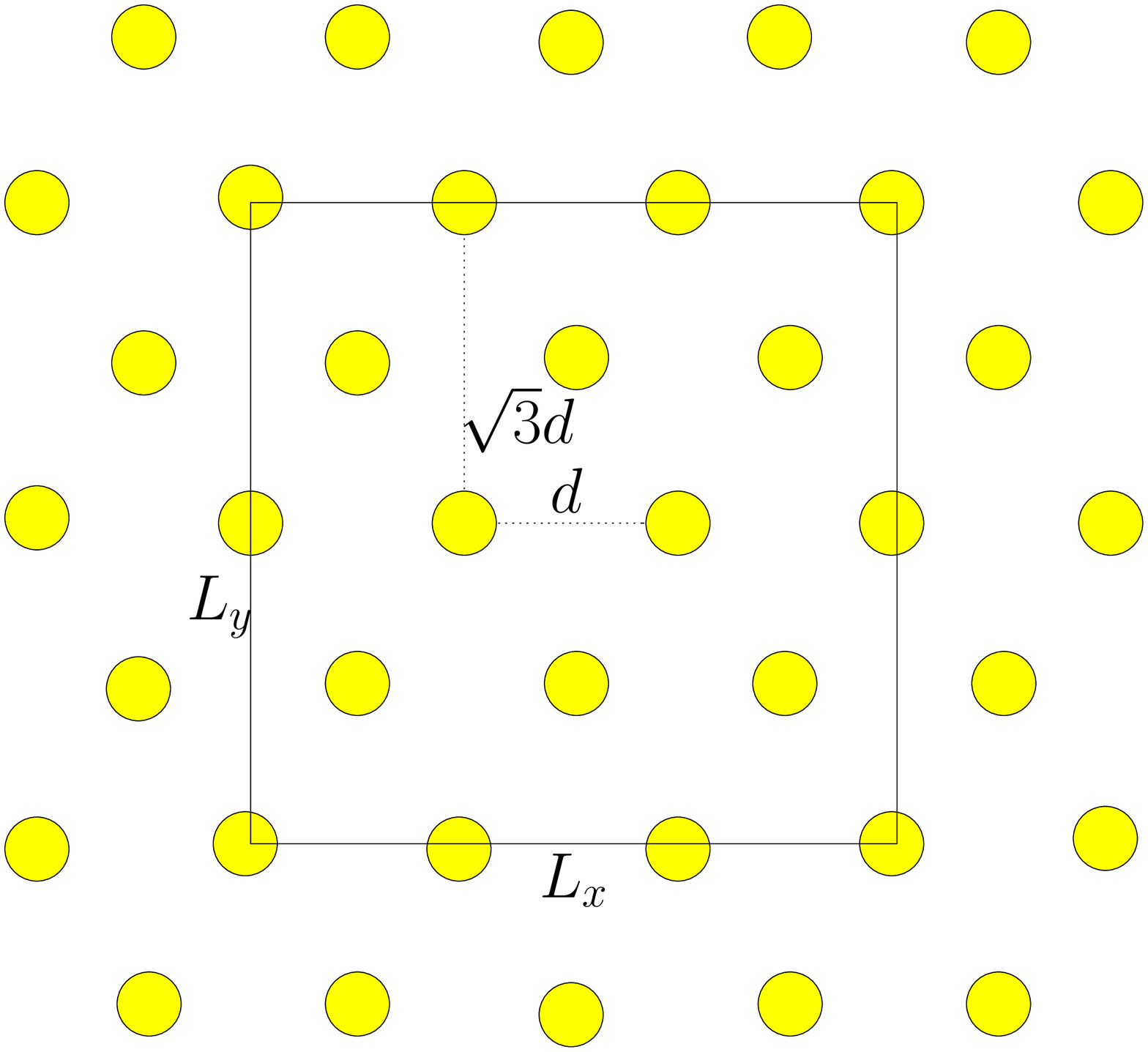}}
\hfill
	\caption{(Color online) 
	(a) A DNA bundle is modeled as a hexagonal lattice
with lattice constant $d$. Individual
DNA molecule is modeled as a hard-core cylinder with negative charges
glued on it according to the positions of nucleotides of a B$-$DNA structure. 
	(b) The periodic simulation box contains 12 DNA cylinders. 
The box dimension in the ($x,y$) plane is shown. The length of the
box in the $z-$axis is three helix periods.
	}
	\label{fig:DNA}
\end{figure}

We assume the DNA molecules in the condensate to arrange in a 
two dimensional hexagonal lattice with lattice constant
$d$ (see Fig. \ref{fig:DNA}a). The DNA axis is parallel to the $z$-axis.
A periodic simulation cell with $N_{DNA} = 12$ DNA molecules 
in the horizontal $(x,y)$ plane and 3 full helix 
periods in the $z$ direction is used (see Fig. \ref{fig:DNA}b). 
Individual DNA molecule
is modeled as an impenetrable cylinder with negative charges glued on it
in accordance with the locations of nucleotide groups
along the double-helix structure of a B$-$DNA. The hardcore cylinder
has radius of 7\AA. The negative charges are hard spheres
of radius 2\AA, charge $-e$ and lie at a 
distance of 9\AA\ from the DNA axis. 
This gives an averaged DNA radius of 1nm. 
The solvent water is treated as
a dielectric medium with dielectric constant $\varepsilon = 78$
and temperature $T=300^oK$. The positions of DNA molecules are fixed in space. 
The mobile ions in solution are modeled as hard spheres with unscreened Coulomb 
interaction 
(the {\em primitive ion} model). For simplicity, all ions have radius of 2\AA.

In practical situation, the DNA bundle is in equilibrium
with an aqueous solution containing free mobile ions at 
given concentrations. Many buffer solutions
contain 50mM of NaCl salt (1:1 salt) and 
10mM of MgCl$_2$ salt (2:1 salt). In this paper, we
focus on the influence of divalent co-ions which is
the case where there is MgSO$_4$ salt (2:2 salt) present
in solution. Overall, our DNA bundle is in equilibrium with 
a solution containing a mixture of as many as three different salts.
Clearly, due to the present of DNA charges,
the concentration of salts inside DNA bundle will be different
than those in the bulk. Mutual correlations among different ions
may favor one particular salt to other.
Thus, to properly simulate the DNA bundle for these
systems, and to properly capture the electrostatic screening
in the system, one needs to go beyond canonical simulation. 
Here, we simulate the system using 
Grand Canonical Monte-Carlo method.
The detail of this method and parameters of our systems, the chemical potentials
of ion species can be found in other works \cite{CohenGCMC,NguyenJCP2016,NguyenSM2016}.
To study the influence of divalent co-ions, two solutions with different salt
mixtures are simulated and compared to each other. 
Solution A contains a mixture of 50mM NaCl and a varying 
concentration of MgCl$_2$ salt. Solution B contains a mixture of
50mM NaCl, 10mM MgCl$_2$, and a varying concentration of MgSO$_4$ salt.
For each simulation run, about 500 million MC moves are 
carried out. To ensure thermalization, about 50 million initial moves 
are discarded before doing statistical analysis of 
the result of the simulation. All simulations are done using the physics simulation library
SimEngine developed by one of the author (TTN).

\section{Result and Discussion\label{sec:discussion}}

\subsection{Monovalent salt differences between the two solutions}
\label{sec:Na}

\begin{figure}[ht]
	\centering
	\resizebox{7cm}{!}{\includegraphics{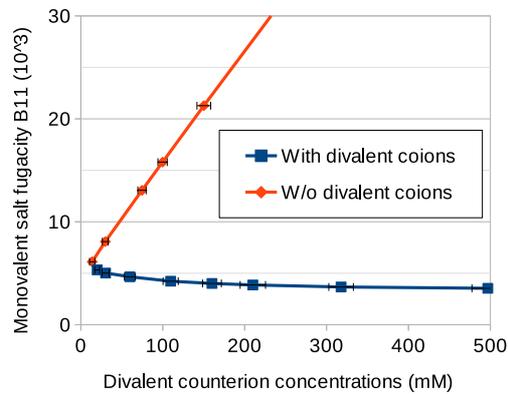}}
	\caption{(Color online) The chemical potential of  a function of 
		divalent counterion concentrations.
		The diamond symbols is for the case of solution A,
		the square symbols is for the case of solution B.
	}
	\label{fig:mu11}
\end{figure}

In Fig. \ref{fig:mu11}, we plot the scaled fugacity of the monovalent salt, 
$B_{1:1} = (V^2/{\Lambda_{+}^{3} \Lambda_{-}^{3}})e^{\beta\mu_{1:1}}$, in a bulk solution
containing 50mM 1:1 salt as a function of divalent counterion concentration in the same
solution. Here $V$ is the volume of the system, $\Lambda_i$ are the thermal wavelengths
of each ion species, and the salt chemical potential is an algebraic sum of 
the chemical potentials of individual ion species
$\mu_{1:1}=\mu_{+}+\mu_{-}$. The volume used for these plots is $V\approx 3300nm^3$.
With increasing counterion concentrations, $c_{2+}$,
the monovalent salt fugacity B$_{1:1}$ increases rapidly for solution type A 
(without divalent co-ions),
while it decreases as much as 40\% for solution type B. 
The observed increase in solution A can be contributed to
the increase in the concentration of the
monovalent Cl$^-$ ions that accompanies the divalent counterions,
Mg$^{+2}$, in solution A. As the divalent counterions concentration
increases, the Cl$^-$ concentration also increases leading to
a higher chemical potential for the 1:1 salt. The linear dependence of the fugacity
on the concentration agrees with this.
The observed decrease in solution B 
can be understood as a consequence of the Gibbs-Duhem theorem
where, in a mixture, increasing the concentration of one species
(keeping the concentration of other species constant), leads
to a reduction in the chemical potential of other species.

\begin{figure}[ht]
	\centering
\hfill
\subfloat[]{\includegraphics[height=2.2in]{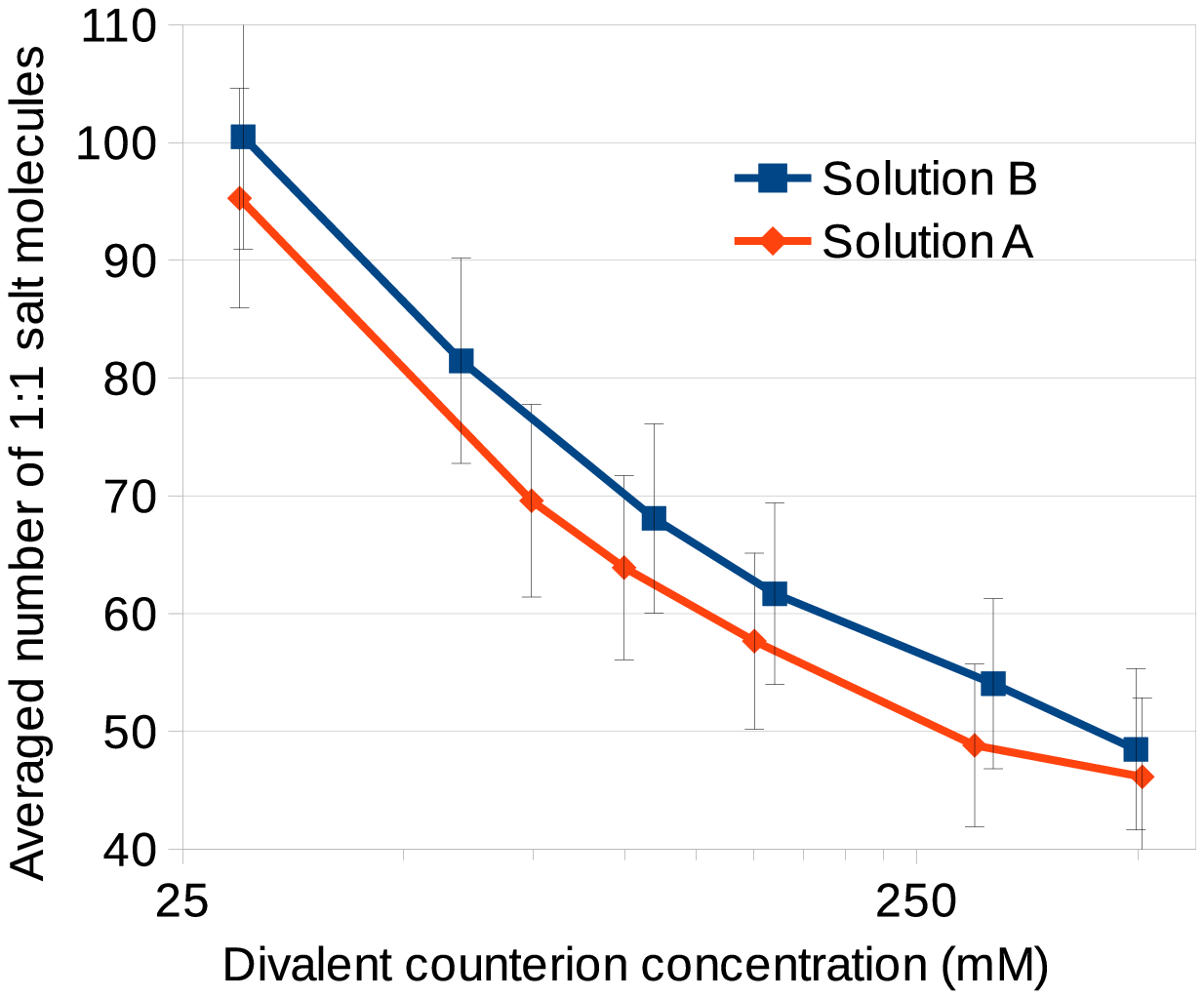}}
\hfill
\subfloat[]{\includegraphics[height=2.2in]{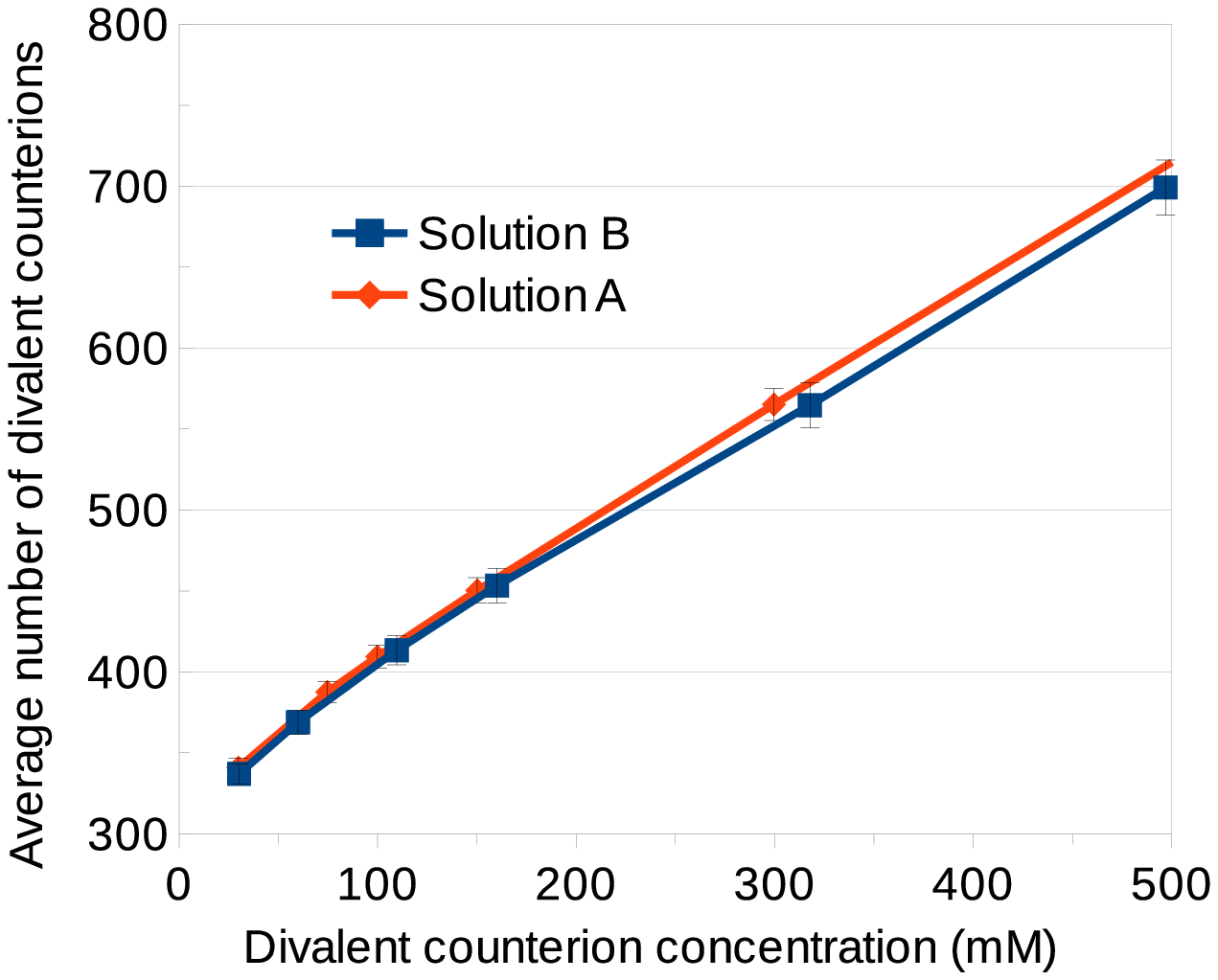}}
\hfill
	\caption{(Color online) The number of 1:1 salt molecules (a) and
	 Mg$^{+2}$ counterions 	(b)
	inside a DNA bundle with $d=40$
	corresponding to a simulation volume of $V_{\mbox{cell}}\approx 1700$ nm$^3$.
	Both are plotted as a function of 
	divalent counterion concentrations.
	The diamond symbols is for the case of solution A,
	the square symbols is for the case with solution B.
	}
	\label{fig:Na}
\end{figure}

Figure \ref{fig:mu11} shows that the fugacity of 1:1 salt for solution B
is smaller than that of solution A at the same divalent counterions.
This suggests that it is easier to insert a monovalent salt molecule 
into the DNA condensate in equilibirum with solution A.
This is indeed true.
In Fig. \ref{fig:Na}a, the averaged total
number of monovalent salt molecules, $n_1$, in the DNA
condensate as a function of the divalent counterion concentrations is plotted.
The data is for the case of inter$-$DNA distance of $d=40$\AA. 
One can see that, in both cases, $n_1$ decreases with increasing counterion concentrations.
This is to be expected. At low concentration $c_{2+}$,
DNA molecules are screened by monovalent
counterions; at high concentration $c_{2+}$, DNA molecules
are screened by divalent counterions. 
This behaviour can only be captured
in a grand-canonical simulation, not in standard canonical simulation
where the number of mobile ions are fixed in advanced.
Importantly, Fig. \ref{fig:Na}a shows that, although the number
are within the error bars, 
the number of monovalent salt molecules is systematically 
higher by as much as 10\%
in the presence of divalent co-ions. 
This result agrees with what was stated earlier that
divalent co-ions make it easier for
monovalent salt to enter the DNA condensate.

To avoid confusion, it should be noted here that in our simulation,
the bulk and the DNA condensate are in thermodynamic equilibrium,
they have the same chemical potentials for component salts.
When we said higher chemical potential, we means that it is higher 
than that of the other solution (chemical potential 1:1 salt 
of solution A is higher than that of solution B).  
In other words, when the chemical potential is high, 
it is costlier to insert the salt molecule to either 
the DNA condensate or the bulk. Vice versa, when the 
chemical potential is lower, it is easier to insert 
the salt molecule into either the bulk or the DNA condensate. 
Fig. \ref{fig:Na}a also supports our claim that more 1:1 salt 
are presence in the DNA condensate in the case of 
divalent co-ions.

For comparison, in Fig. \ref{fig:Na}b, the number of Mg$^{+2}$ counterions is
also plotted for different concentrations. The number of Mg$^{+2}$
needed to neutralize the DNA is 360 ions. At low concentration,
the number of Mg$^{+2}$ is slighly below this number suggesting
that DNA is also screened by monovalent counterions. At high concentration
the number of Mg$^{+2}$ ions is more than needed to neutralize
the DNA and the number of 1:1 salt is reduced by 60\%
at the highest concentration simulated. The system is dominated
by the divalent ions.
This is in agreement with the grand-canonical
picture that we use.
Note that, we are never in a counterion-only
regime even at low divalent salt concentration. In this
limit, the DNA should be screened by monovalent salt present
in solution. 

\subsection{Overcharging of DNA at high divalent counterion concentration}

Next, let us look at the overcharging of DNA by multivalent
counterions at high concentration, $c_{2+}$.
\begin{figure}[ht]
	\centering
\hfill
\subfloat[]{\includegraphics[height=2.3in]{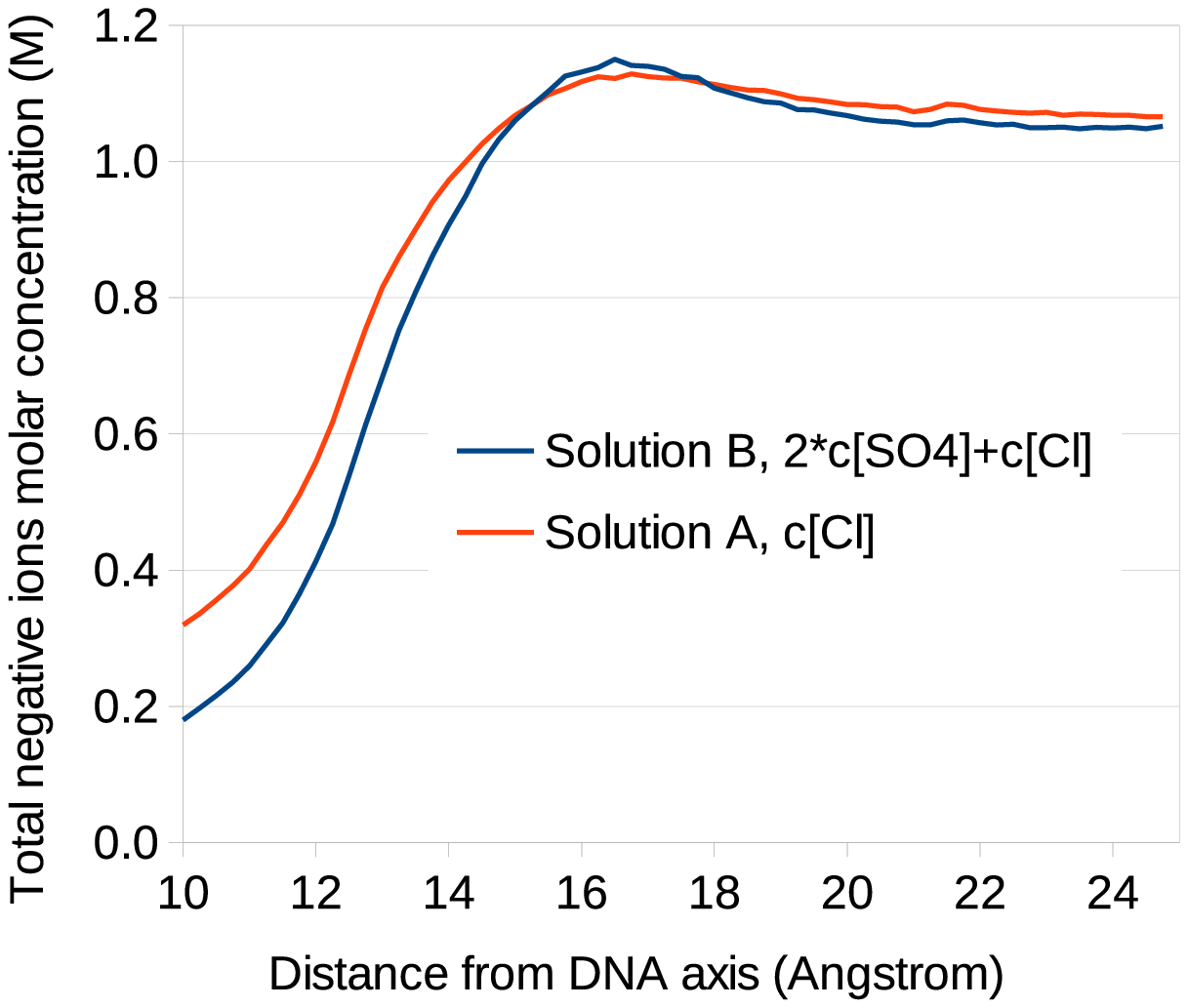}}
\hfill
\subfloat[]{\includegraphics[height=2.2in]{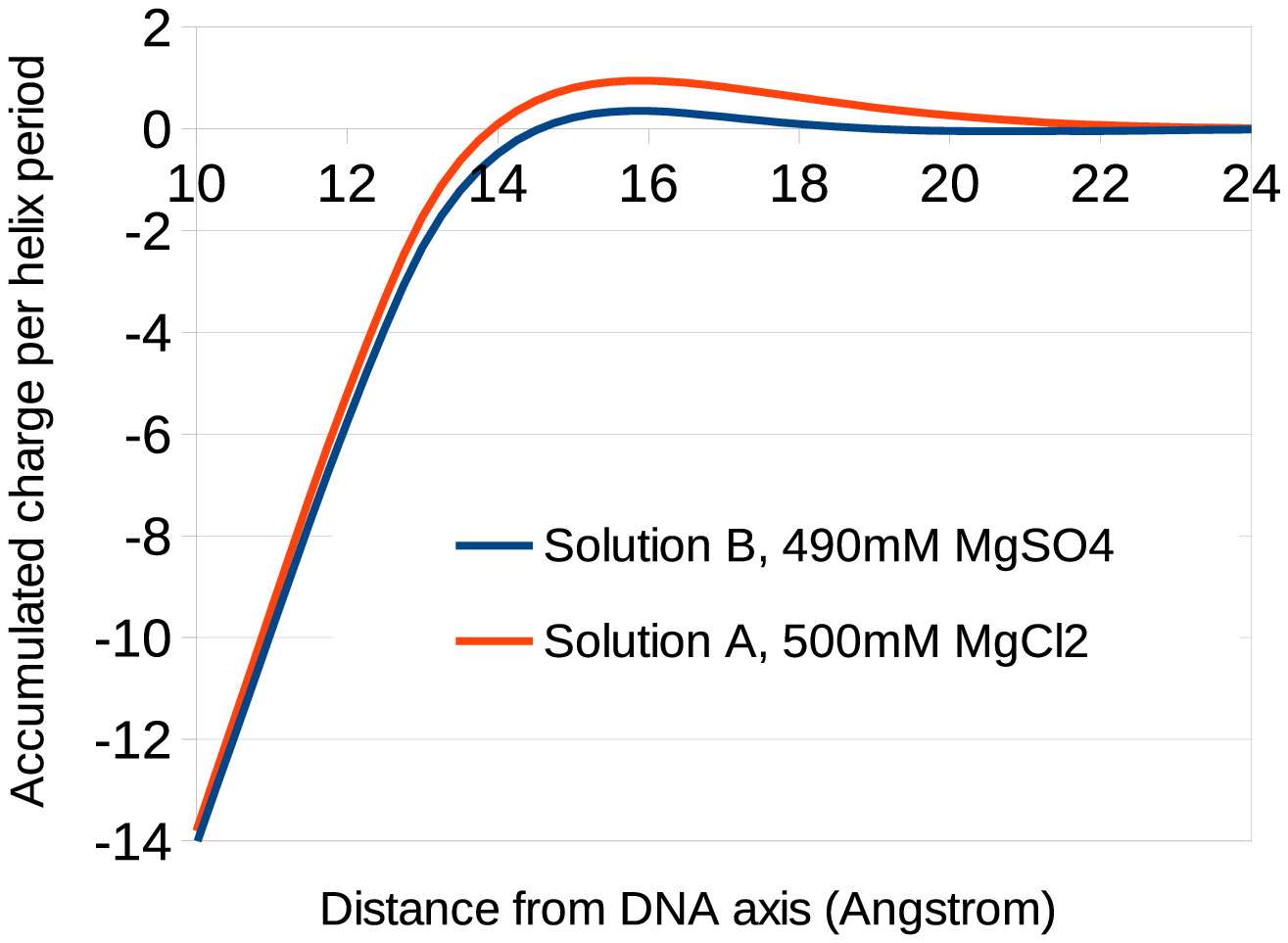}}
\hfill
	\caption{(Color online) (a) The total charge density of negative ions
	as a function of distance from DNA axis, in unit of molar concentration.
b) The accumulated amount of charges per helix DNA from DNA surface in units
of elementary charges $e$. It is $-20e$ at the DNA surface (not shown).}
	\label{fig:rhom}
\end{figure}
In Fig. \ref{fig:rhom}a we plot total charge density of the negative ions 
as a function of distance from a DNA cylinder. 
For solution A, 
this is simply the charge density of Cl$^{-}$ co-ions, $-ec_{\mbox{Cl}}$.
For  solution B, 
this is the sum $-c_{\mbox{Cl}}-2ec_{\mbox{SO4}}$. 
These density profiles are plotted for the divalent counterion concentration
of $c_{2+}\approx 500mM$ and for inter$-$DNA distance
of $d=50$\AA. 
One can see that at large distance $r$ from DNA axis,
$r\geq 17$\AA, the negative charge density {\em decreases}
with increasing $r$. 
This means, for an observer from far away, the 
negative ions are accumulated near the DNA or
the apparent net charge of DNA is {\em positive}.
Indeed, the charge density should obey the Boltzmann distribution 
and is higher at the place of lower electrostatic potential energy. 
In Fig. \ref{fig:rhom}, for negative charges, 
their concentration {\em increases} as one approaches 
the DNA from far away. Since the electrostatic potential at infinity 
is zero, this increase means that the electrostatic potential 
is becoming more positive as one approaches the DNA. 
In other words, DNA apparent charge is positive. 
This is a clear indication of an overcharging effect. 

Another conclusion one can draw from Fig. \ref{fig:rhom}a is
that from distance $r=17$\AA\ down to $r=10$\AA,
the negative charge density decreases with decreasing $r$. 
One can deduce that this range of $r$ corresponds the
condense layer of counterions around DNA. 
Our result shows that, the length of this counterion condensation layer 
is approximately 7\AA\ within the surface of a DNA molecule.
This agrees with the fact that if Mg$^{2+}$ counterions
were to form a two dimensional strongly correlated liquid on
the DNA surface, the averaged distance between neighbor counterions
in this liquid is also about 7\AA. Additionally, the discreteness of DNA
charges is also about 7\AA.

The overcharging effect is more evident if one plots the accumulation
of all charges, positive and negative, from the DNA axis. 
In Fig. \ref{fig:rhom}b
the accumulated charge (DNA and all ions) as function of distance from
a DNA's axis is plotted for one helix period. At the DNA surface,
the amount of charges per period is $-20e$. As one moves further
away from the surface, positive counterions condense on DNA 
to reduce its charges. At the distance of $r\approx 14-16$\AA, the
accumulated charge starts to become positive and it is
maximally positive at $r \approx 16$\AA. This once again
confirms the overcharging of DNA.

Fig. \ref{fig:rhom}a and Fig. \ref{fig:rhom}b
 also show that the presence of divalent co$-$ions have
strong quantitative influence on DNA overcharging degree.
Within the condensation layer, 
the negative divalent co-ions are repelled from DNA much stronger in the
presence of divalent co-ions compared to the case without divalent co-ions.
This is to be expected. For solution B at
high divalent co-ion concentration, the negative ions are mostly made
of divalent co-ions. The repulsion
from DNA negative charges on the co$-$ions is larger in solution
B than that of solution A. Hence less negative charges in the condensation
layer around each DNA.

Outside this layer, $r\geq 17$\AA, the peak of the total negative ion
concentration for solution B 
(the blue curve) is slightly higher. The slope of this curve is also higher. 
This mean that negative ions are accumulated at the surface of the
condensed layer. This leads to a reduction in the correlation
among divalent counterions, hence a significant reduction in overcharging
degree as shown in Fig. \ref{fig:rhom}b.

\subsection{DNA$-$DNA ``effective" interaction mediated by counterions}

\begin{figure}[ht]
	\centering
\hfill
\subfloat[]{\includegraphics[height=2.2in]{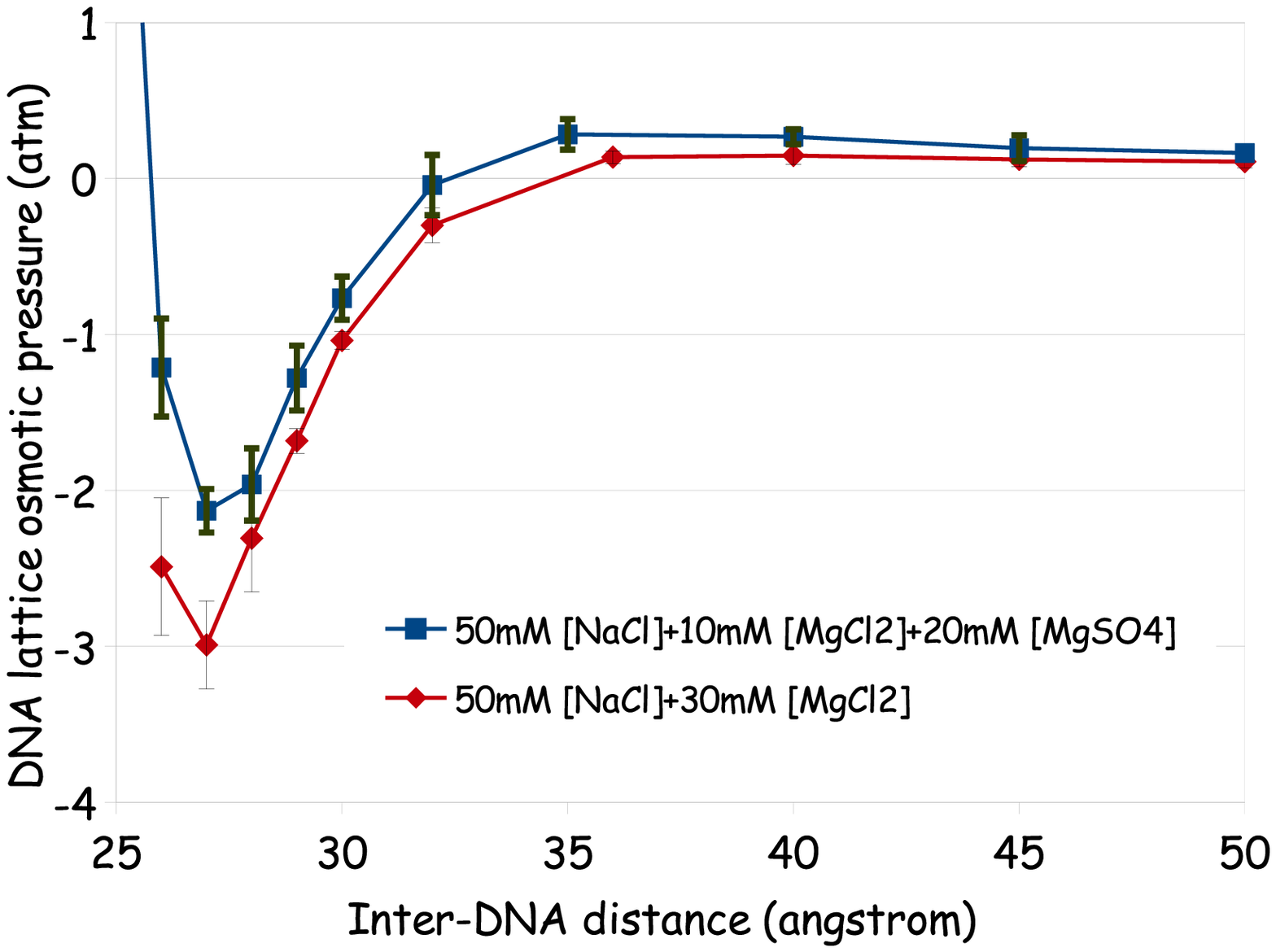}}
\hfill
\subfloat[]{\includegraphics[height=2.2in]{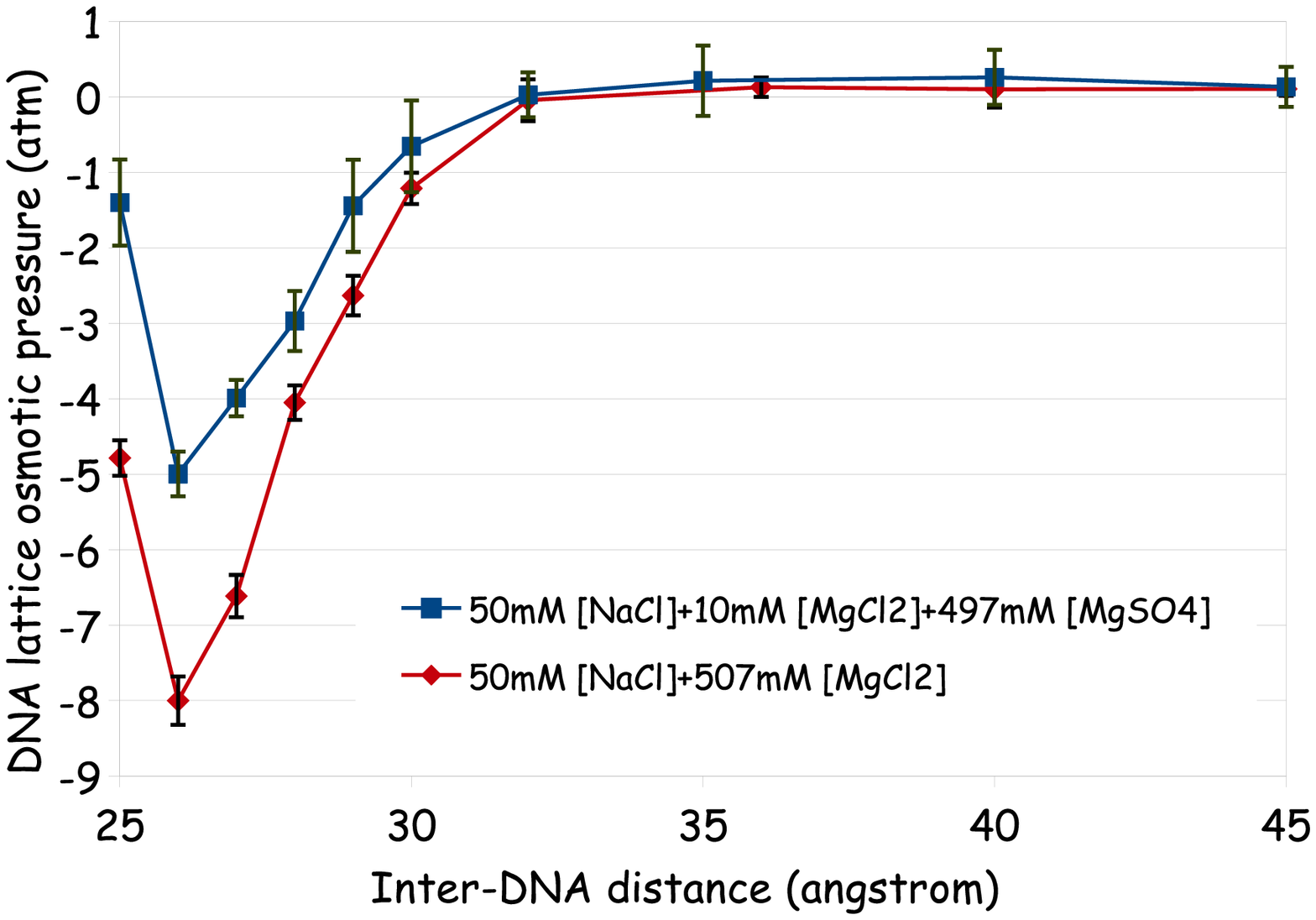}}
\hfill
	\caption{(Color online) The osmotic pressure of the DNA bundle
		as function of the interaxial DNA distance $d$ for 
		divalent counterion concentration of (a) 30mM $-$ low concentration; 
and (b) 507mM $-$ high concentration.
		The diamond symbols are for the case of no divalent co-ions 
		(solution A),
		the square symbols are for the case with divalent co-ions
		(solution B)
		in bulk the solution. Explicit values for the concentration
of component salts are given in the corresponding figure legend.
		The solid lines are guides to the eye. 
	}
	\label{fig:posm20}
\end{figure}

In the presence of divalent coions, the ionic strength of the solution
is larger. Additionally, as section \ref{sec:Na} shows, more
the monovalent salt molecules enter the condensate in the
case of solution B. These factors lead to a 
smaller Debye screening radius, $r_s$, for solution B.
This weakens electrostatic interactions among DNA molecules. 
This is indeed what we observed in simulation. 
Fig. \ref{fig:posm20}a plots the osmotic pressure of DNA bundle 
as a function of the inter$-$axial DNA distance, $d$, 
for solution A containing about 30mM MgCl$_2$ salt concentration
and solution B containing about 20mM MgSO$_4$ salt concentration. 
Both solution contains divalent counterion concentration of 
$c_{2+}\simeq 30mM$. Similarly, Fig. \ref{fig:posm20}b plots
the osmotic pressure of DNA bundle as a function of the inter$-$axial DNA 
distance for the case of high divalent counterion concentration
of $c_{2+}\simeq 500mM$.
Because this osmotic pressure is directly related to
the ``effective'' force between DNA molecules at that inter$-$axial
distance \cite{Nordenskiold95,NordenskioldJCP86}, 
this figure also serves as a plot of DNA$-$DNA effective interaction
mediated by multivalent counterions.
In all cases, there is a short$-$range attraction between two DNA molecules as 
they approach each other. This is the well-known phenomenon of like$-$charge 
attraction
between macroions \cite{NetzLikeChargedRods,GelbartPhysToday,NguyenRMP2002}. 
The maximum attraction 
occurs at the distance $d\simeq 26-27$\AA, in good agreement with various 
theoretical and 
experimental results \cite{Parsegian92,Phillips05}. 
For smaller $d$, the DNA-DNA interaction
experiences sharp increase due to the hardcore repulsion
between the counterions. 

From Fig. \ref{fig:posm20}, it can be seen that the DNA$-$DNA attraction at short
range ($r < 30\AA$) is reduced in the present of divalent co$-$ions at both
low concentration and high concentration of the divalent counterions. The overall
reduction can be as much as 40\% in the presence of divalent co$-$ions.

On the other hand, at larger distance, the DNA$-$DNA repulsive interaction
among undercharged (low $c_{2+}$) and overcharged (high $c_{2+}$)
is stronger in solution B. This means that divalent co$-$ions donot play
the simple role of linear Debye$-$H\"{u}ckel screening of the "dressed"
DNA charge. Rather, they bind to divalent counterions and reduce
the amount of charges available to participate in linear screening as
the dressed counterion theory suggested \cite{NajiDressIon2010}.

\begin{figure}[ht]
	\centering
	\resizebox{9cm}{!}{\includegraphics{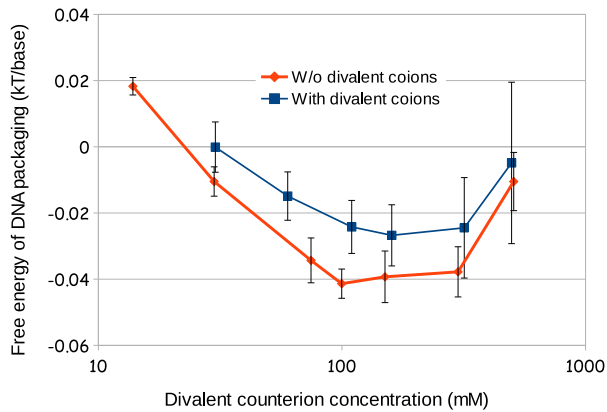}}
	\caption{(Color online) The free energy of packaging DNA molecules
		into condensate as a function of the divalent counterion concentrations.
	}
	\label{fig:FvC}
\end{figure}

From the P-V curves such as those in Fig. \ref{fig:posm20}a and Fig. \ref{fig:posm20}b,
one can calculate the free energy of packaging DNA into condensate by 
integrating the pressure with the volume.
This 
free energy is plotted in Fig. 
\ref{fig:FvC} as function of the divalent counterion concentrations.
It shows a non-monotonic dependence of the electrostatic contribution to DNA 
packaging free energy.
This is common for DNA condensation by multivalent counterions
and is the result of interplay between DNA overcharging effect
and DNA$-$DNA like charge attraction. It is consistent with 
the correlation theory of DNA reentrant condensation 
by multivalent counterions \cite{Shklovskii1999,NguyenRMP2002,NguyenJCP2000}
and the experiment results on ejecting DNA from bacteriophage under varying
counterion concentrations \cite{Knobler08}. At low (high) $c_{2+}$, DNA molecules
is undercharged (overcharged) and repel each other. At medium $c_{2+}$,
DNA like-charge attractions cause to condensate into bundles.
Fig. \ref{fig:FvC} further shows that the DNA packaging free energy
is higher in the presence of divalent co-ions. This can be 
understood as the consequence of all the physics presented so far: reduction
of DNA$-$DNA like$-$charge interaction and overcharging.
This also agrees with the experimental fact that MgSO$_4$ cause
earlier DNA decondensation, and more DNA ejected from
bacteriophages\cite{Knobler08} compared to MgCl$_2$.

\section{Conclusion\label{sec:conclusion}}

In conclusion, a computational study of the influence of multivalent co-ions
on strongly correlated electrostatics of
DNA condensation by multivalent counterions is presented.
Divalent coions' influence is multiple folds. First,
in a grand canonical equilibrium with particle reservoir
at given ion concentrations, divalent co-ions reduce the cost of adding 
monovalent salt to the DNA bundle system by as much as 
40\% in terms of fugacity, 10\% in terms of concentration. 
Because monovalent salts mostly
participate in screening of electrostatic interaction in the system,
more monovalent salt enter the bundle leads to screening out of
short-range DNA$-$DNA like charge attraction and weaker DNA condensation free energy.
Secondly, the strong repulsion between DNA and divalent co-ions
and among divalent co-ions leads to depletion of negative ions
near DNA surface as compared to the case without divalent
co$-$ions. The condensation layer of counterions near DNA surface
is shown to be about 7\AA. In the presence of divalent co$-$ions,
negative charge concentration is slightly higher
at the surface of the condensed layer. This also leads to 
significant reduction of DNA overcharging. The overall results
of our study is the reduction of DNA strongly correlated
electrostatics and agrees well with experimental
fact that MgSO$_4$ solution causes stronger DNA ejection from
bacteriophages than MgCl$_2$ solution.

In our computer simulation, we focus on the specific case of 
divalent coions to compare with monovalent counterions. 
We believe most important qualitatively behaviours 
will not change significantly for higher counterion, coion valency. 
Additionally, there are many other facotrs that were neglected
in our simplified model. One approximation is that in the simulation,
the position of the DNA cylinders are straight with 
infinite bending rigidity. Inside viruses, 
DNA are bent, and the configuration entropy of the DNA are not necessary zero, and there is
not a perfect hexagonal arrangement of DNA cylinder with fixed inter-DNA distance
\cite{Phillips05}.
The relative orientation of individual DNA cylinder is also fixed
in our simulation. Allowing for various orientation of DNA charges
can lead to frustration effect similar to spin glass in two dimension
Ising model \cite{GrasonPRL2010}.
The physical parameters of the system such as ion sizes
can also affect the strength of DNA$-$DNA short range attraction
\cite{NguyenJCP2016}. 
The dielectric discontinuity at the DNA surface is also
another important factor that was neglected in this model
and can have important consequence on the screening of DNA
in different electrolytes \cite{NajiJCP2014}.
Inclusion of all these factors is beyond the scope of this
paper and need to be considered in more detail in a future
work. Nevertheless, we believe that the qualitative difference
among DNA condensation with monovalent coion and with divalent
coions discussed in this paper will not be affected.


%

\begin{acknowledgements}
We would like to thank Drs. A. Lyubartsev, B. Shklovskii, A. Evilevich,
Tung Le, T. X. Hoang for valuable discussions. 
TTN acknowledges the financial support of
the Vietnam National Foundation for Science and Technology
NAFOSTED Contract 103.02-2012.75. The authors are indebted to Dr
A. Lyubartsev for providing us with the 
Fortran source code of their Expanded Ensemble Method for
calculation of osmotic pressure.

\end{acknowledgements}


\end{document}